\documentclass[oldversion]{aa}
\usepackage{graphicx}
\def\H0 {$H_{\rm o}$}

\def\CH3C2H {\hbox{${\rm CH}_3{\rm C}_2{\rm H}$}} 

\def\ffas {\hbox{$\,.\!\!^{\prime\prime}$}}

\def\ffm {\hbox{$\,.\!\!^{\rm m}$}}
\def\ffs {\hbox{$\,.\!\!^{\rm s}$}}
\def \la{\mathrel{\mathchoice   {\vcenter{\offinterlineskip\halign{\hfil
$\displaystyle##$\hfil\cr<\cr\sim\cr}}}
{\vcenter{\offinterlineskip\halign{\hfil$\textstyle##$\hfil\cr
<\cr\sim\cr}}}
{\vcenter{\offinterlineskip\halign{\hfil$\scriptstyle##$\hfil\cr
<\cr\sim\cr}}}
{\vcenter{\offinterlineskip\halign{\hfil$\scriptscriptstyle##$\hfil\cr
<\cr\sim\cr}}}}}
\def \ga{\mathrel{\mathchoice   {\vcenter{\offinterlineskip\halign{\hfil
$\displaystyle##$\hfil\cr>\cr\sim\cr}}}
{\vcenter{\offinterlineskip\halign{\hfil$\textstyle##$\hfil\cr
>\cr\sim\cr}}}
{\vcenter{\offinterlineskip\halign{\hfil$\scriptstyle##$\hfil\cr
>\cr\sim\cr}}}
{\vcenter{\offinterlineskip\halign{\hfil$\scriptscriptstyle##$\hfil\cr
>\cr\sim\cr}}}}}

\begin{document}

\title{36\,GHz methanol lines from nearby galaxies: maser or 
       quasi-thermal emission?}

\author{P.~K. Humire \inst{1},
        C. Henkel \inst{1,2,3},
        Y. Gong \inst{1},
        S. Leurini \inst{4,1},
        R. Mauersberger \inst{1},
        S.~A. Levshakov \inst{5,6,7},
        B. Winkel \inst{1},
        A. Tarchi \inst{4},
        P. Castangia \inst{4},
        A. Malawi \inst{2},
        H. Asiri \inst{2},
        S.~P. Ellingsen \inst{8},
        T.~P. McCarthy \inst{8,9},
        X. Chen \inst{10,11},
        X. Tang \inst{3,12}
       }

\offprints{C. Henkel, \email{chenkel@mpifr-bonn.mpg.de}}

\institute{
   Max-Planck-Institut f{\"u}r Radioastronomie, Auf dem H{\"u}gel 69, 
   53121 Bonn, Germany
  \and
   Astronomy Department, Faculty of Science, King Abdulaziz University, P.O. Box 80203,
   Jeddah 21589, Saudi Arabia
  \and
   Xinjiang Astronomical Observatory, Chinese Academy of Sciences, 830011 Urumqi, China
  \and
   INAF-Osservatorio Astronomico di Cagliari, Via della Scienza 5, 09047, Selargius (CA), Italy
  \and
   Ioffe Physical-Technical Institute, 194021, St. Petersburg, Russia
  \and
   Electrotechnical University `LETI', 197376 St.Petersburg, Russia
  \and
   Petersburg Nuclear Physics Institute, 188300 Gatchina, Russia
  \and
   School of Natural Sciences, University of Tasmania, Hobart, TAS 7001, Australia
  \and
   Australia Telescope National Facility, CSIRO, PO Box 76, Epping, NSW 1710, Australia
  \and
   Center for Astrophysics, GuangZhou University, Guangzhou 510006, China
  \and
   Shanghai Astronomical Observatory, Chinese Academy of Sciences, Shanghai 200030, China
  \and
   Key Laboratory of Radio Astronomy, Chinese Academy of Sciences, 830011 Urumqi, China}

\date{Received date ; accepted date}
 
\abstract
{
Methanol (CH$_3$OH) is one of the most abundant interstellar molecules, 
offering a vast number of transitions to be studied, including many maser 
lines. However, while the strongest Galactic CH$_3$OH lines, the so-called
class~II masers, show no indications for the presence of superluminous 
counterparts in external galaxies, the less luminous Galactic class I sources
appear to be different. Here we report class~I 36\,GHz ($\lambda$ $\approx$ 0.8\,cm)
CH$_3$OH 4$_{-1}$ $\rightarrow$ 3$_0$~E line emission from the nearby 
galaxies Maffei~2 ($D$ $\approx$ 6\,Mpc) and IC~342 ($D$ $\approx$ 3.5\,Mpc), 
measured with the 100m telescope at Effelsberg at three different epochs 
within a time span of about five weeks. The 36\,GHz methanol line of 
Maffei~2 is the second most luminous among the sources detected with certainty 
outside the Local Group of galaxies. This is not matched by the moderate infrared 
luminosity of Maffei~2. Higher-resolution data are required to check whether this is related 
to its prominent bar and associated shocks. Upper limits for M~82, NGC~4388, NGC~5728
and Arp~220 are also presented. The previously reported detection of 36\,GHz maser 
emission in Arp~220 is not confirmed. Nondetections are reported from the related 
class~I 44\,GHz ($\lambda$ $\approx$ 0.7\,cm) methanol transition towards Maffei~2 
and IC~342, indicating that this line is not stronger than its 36\,GHz counterpart.
In contrast to the previously detected 36\,GHz CH$_3$OH emission in NGC~253 
and NGC~4945, our 36\,GHz profiles towards Maffei~2 and IC~342 are
similar to those of previously detected nonmasing lines from other molecular 
species. However, by analogy to our Galactic center region, it may well be possible 
that the 36\,GHz methanol lines in Maffei~2 and IC~342 are composed of 
a large number of faint and narrow maser features that remain 
spatially unresolved. In view of this, a search for a weak broad 36\,GHz line 
component would also be desirable in NGC~253 and NGC~4945.} 


\titlerunning{Methanol in galaxies}

\authorrunning{Authors}

\maketitle

\section{Introduction}

Methanol (CH$_3$OH) is one of the most abundant interstellar molecules (e.g., 
Kalenski \& Sobolev 1994; Wang et al. 2004; Maffucci et al. 2018) exhibiting 
a plethora of lines at centimeter(cm)-, millimeter(mm)-, and submm  wavelengths. 
To provide an example, Schilke et al. (2001) and Comito et al. (2005) reported 
the detection of a total of $\approx$650 methanol transitions from Orion-KL. 
Some of these lines show inverted level populations and significant optical 
depths. Such methanol maser lines form two distinct families exhibiting 
clearly nonthermal emission in quite specific but different
transitions: class~I masers are often separated from the main sources of 
star forming activity (between 0.1 and 1.0\,pc from UC~H{\sc ii} regions, OH
and H$_2$O masers according to G{\'o}mez-Ruiz et al. 2016), whereas class~II 
masers are closely associated with such sites. The 6.7\,GHz transition 
(Menten 1991) has become the prototypical class~II maser line, exhibiting 
particularly strong emission in regions of intense radiation from warm dust 
and relatively cool gas (Cragg et al. 2005). Sources 
characterized by class~I emission are encountered in regions not necessarily 
devoid of far infrared radiation but requiring collisional excitation -- 
they appear to be associated with weak shocks, possibly related to outflows 
from young stellar objects interacting with the dense ambient interstellar 
medium (e.g., Kurtz et al. 2004; Leurini et al. 2016). In such regions, 
the 4$_{-1}$ $\rightarrow$ 3$_0$ E and 7$_0$ $\rightarrow$ 6$_1$ A$^+$ 
transitions at 36 and 44\,GHz, respectively, become most prominent. 
Apparently, the 36\,GHz line is strongest towards sources with kinetic 
temperatures of $T_{\rm kin}$ $\la$ 100\,K, where signs of high-mass
star formation are not yet seen. The 44\,GHz masers are closer 
to ultracompact (UC)-H{\sc ii}-regions and mm wave continuum 
sources and may be excited at $T_{\rm kin}$ $\ga$ 100\,K. With respect 
to required densities, optimal conditions for 36\,GHz masers appear to 
include densities that are an order of magnitude higher than those at 44\,GHz 
(e.g., Pratap et al. 2008; McEwen et al. 2014; Nesterenok 2016; Leurini 
et al. 2016). 

More than one thousand methanol masers are known in the Galaxy 
(e.g., Menten 1991; Pratap et al. 2008; Yusef-Zadeh et al. 2013; Cotton
\& Yusef-Zadeh 2016; Yang et al. 2017, 2019). However, 
in extragalactic targets, such maser lines are rarely 
observed. Following the detections of quasi-thermal methanol emission 
in nearby galaxies (Henkel et al. 1987), the large number of Galactic 
masers and the existence of even more luminous H$_2$O and OH 
``megamasers'' (e.g., Lo 2005) provided strong motivation to 
search for class~II 6.7\,GHz maser emission towards extragalactic 
sources (Ellingsen et al. 1994a, Phillips et al. 1998; Darling et 
al. 2003). This yielded detections in absorption towards NGC~3079 
and Arp~220 (Impellizzeri et al. 2008; Salter et al. 2008) and 
in emission towards the Large Magellanic Cloud (LMC; Sinclair et 
al. 1992; Ellingsen et al. 1994b; Beasley et al. 1996; Green et al.
2008; Henkel et al. 2018a) and the Andromeda galaxy M~31 (Sjouwerman
et al. 2010). The LMC also provided a 12.2\,GHz class~II maser 
detection, reported by Ellingsen et al. (2010). However, the intrinsic 
brightness of all these emission lines turned out to be similar 
to those of their stronger Galactic counterparts.

Since none of the surveys for ultraluminous class~II masers revealed 
positive results, searches for brighter analogs of the relatively 
inconspicuous Galactic class~I masers were also carried out in nearby 
extragalactic sources. The detection of 36\,GHz class~I masers in 
NGC~253 by Ellingsen et al. (2014) revealed emission that is more than 
ten times more luminous than the widespread emission associated with the 300\,pc 
$\times$ 100\,pc central region of our Galaxy and 10$^4$ times 
more luminous than a typical individual Galactic 36\,GHz maser. This 
demonstrates the existence of ultraluminous methanol masers (with respect 
to their Galactic counterparts). Follow-up observations confirmed this 
detection (Ellingsen et al. 2017; Gorski et al. 2017; Chen et al. 
2018; Gorski et al. 2019), which was augmented by successful
searches for the related class~I maser transitions at 44 and 84\,GHz 
in NGC~253 (Ellingsen et al. 2017; McCarthy et al. 2018a). Nevertheless, 
the search for additional sources turned out to be difficult. For 
some time, NGC~4945 was the only additional galaxy detected in 
the 36\,GHz CH$_3$OH class~I maser line beyond any reasonable doubt 
(at multiple epochs with more than one telescope; see McCarthy 
et al. 2017, 2018b). More recently, 36\,GHz emission was also reported 
from IC~342 and NGC~6946 (Gorski et al. 2018). Chen \& Ellingsen 
(2018) very briefly mentioned a Jansky Very Long Baseline Array (JVLA) 
detection of Maffei~2. In the following we report clear single-dish 
detections of Maffei~2, a particularly luminous emitter of 36\,GHz line 
radiation, and IC~342 in the 36.169261\,GHz (Endres et al. 2016) 
4$_{-1,4}$ $\rightarrow$ 3$_{0,3}$ E transition of methanol, thus providing 
additional strong evidence for the widespread existence of such emission 
in nearby galaxies. This report also includes a number of nondetections,
most notably the nondetection of Arp\,220.

\begin{figure}[ht]
\hspace{0.3cm}
\resizebox{12.0cm}{!}{\rotatebox[origin=br]{-90}{\includegraphics{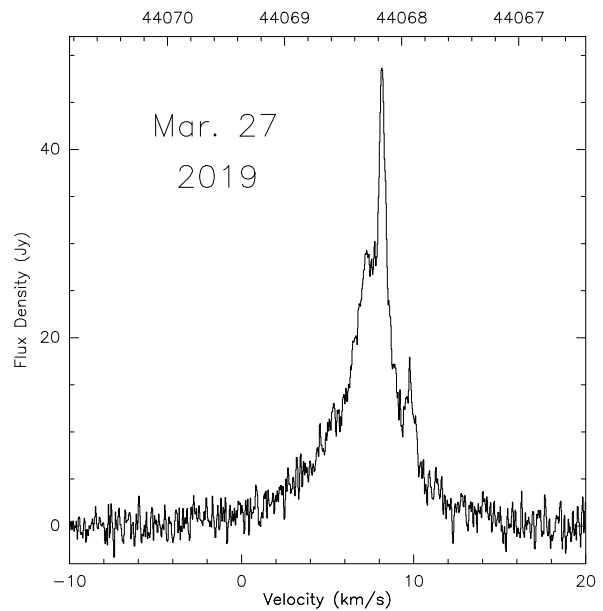}}}
\caption{7$_{0}$ $\rightarrow$ 6$_1$ A$^+$ methanol spectrum from Orion-KL. The adopted
rest frequency is 44.069367\,GHz, the beam size is 23$''$ and the coordinates are 
$\alpha_{J2000}$ = 5$^{\rm h}$ 35$^{\rm m}$ 14\ffs2, $\delta_{J2000}$ = --5$^{\circ}$ 
22$'$ 30$''$. The velocity scale is the local standard of rest and the channel spacing is
0.31\,km\,s$^{-1}$ after averaging eight contiguous channels.} 
\label{orion}
\end{figure}

\section{Observations}

\subsection{Front- and backends}

The 36\,GHz measurements were carried out with the 100m telescope at 
Effelsberg\footnote{This publication is based on observations with the
100m telescope of the MPIfR (Max-Planck-Institut f{\"u}r Radioastronomie)
at Effelsberg.} near Bonn, Germany, in early January and early February
2019. Complementary 44\,GHz observations were taken in late March
2019. We obtained the data with a secondary focus Q-band receiver, 
sensitive in the range 33 -- 50\,GHz, providing both circular polarizations
with an equivalent system temperature of $\approx$130\,Jy on a flux density
scale ($\approx$1.6\,K main beam brightness temperature per Jy) prior to 
averaging the two polarizations. Pointing observations were carried out
every $\approx$40 minutes towards 3C\,84 and NRAO\,150 near Maffei~2 and 
towards W3(OH) near IC~342. The pointing accuracy was 4$''$$\pm$2$''$
(the error is the standard deviation of an individual measurement)
and never exceeded 10$''$, while the beam size was $\approx$23$''$ at 
both frequencies. 

Near the start of each 36\,GHz observing session, we targeted the 
well-known HC$_3$N $J$ = 4$\rightarrow$3 (36.392\,GHz) and SiS $J$ = 
2$\rightarrow$1 (36.310\,GHz) lines of IRC+10216 to make 
sure the system was working. At 44.069\,GHz, we chose the CH$_3$OH 7$_0$ 
$\rightarrow$ 6$_1$~A$^+$ line towards Orion-KL (Fig.~\ref{orion}) as a 
system check that may be used for future projects focusing on issues 
related to maser variability. For the measurements towards the targets 
themselves we used a position-switching mode, with 2~minutes off- and 
on-source integration times, respectively, and with offsets in right 
ascension alternating between \hbox{+10$'$} and \hbox{--10$'$} to remove 
elevation-dependent effects. We employed a Fast Fourier Transform spectrometer 
backend with a bandwidth of 300\,MHz. The number of channels was 65536, yielding a 
channel spacing of 0.038\,km\,s$^{-1}$ at 36\,GHz and a velocity resolution 
of 0.044\,km\,s$^{-1}$ (see Klein et al. 2012). At 44\,GHz, the corresponding 
values are 0.031\,km\,s$^{-1}$ and 0.036\,km\,s$^{-1}$.

\begin{figure*}[ht]
\hspace{0.3cm}
\resizebox{28.0cm}{!}{\rotatebox[origin=br]{-90}{\includegraphics{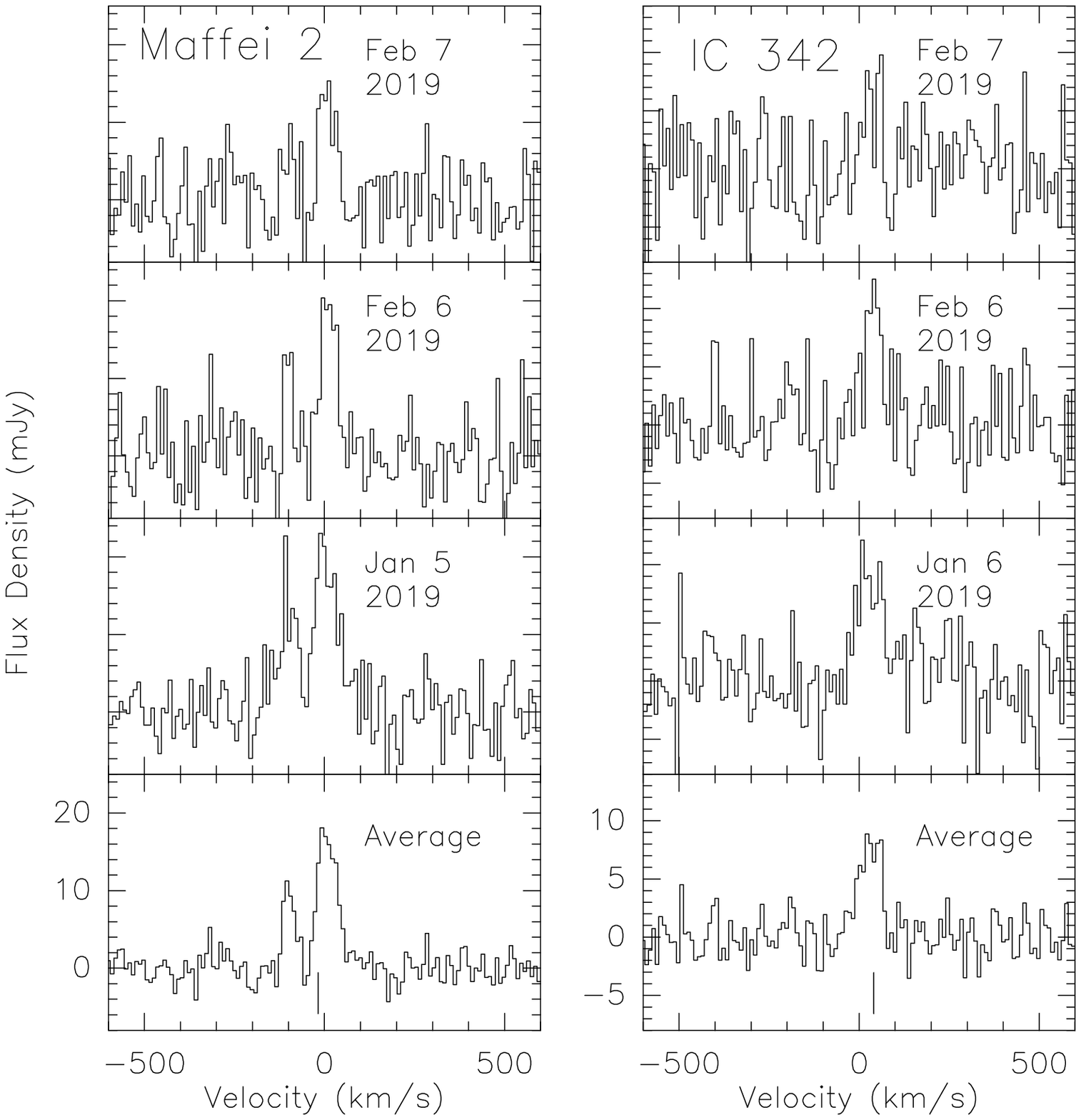}}}
\vspace{-1.0cm}
\caption{4$_{-1}$ $\rightarrow$ 3$_0$ E methanol spectra from Maffei~2 (left)
and IC~342 (right). Spectra from a given source are all presented with the same 
amplitude and velocity scale. The adopted rest frequency is 36.169261\,GHz, the 
coordinates are $\alpha_{J2000}$ = 2$^{\rm h}$ 41$^{\rm m}$ 55\ffs2, $\delta_{J2000}$ 
= 59$^{\circ}$ 36$'$ 12$''$ (Maffei~2) and $\alpha_{J2000}$ = 3$^{\rm h}$ 
46$^{\rm m}$ 48\ffs6, $\delta_{J2000}$ = 68$^{\circ}$ 05$'$ 46$''$ (IC~342). 
The velocity scale is the local standard of rest. The spectra were smoothed by 
averaging 256 contiguous channels to a channel spacing of 9.7\,km\,s$^{-1}$. The 
average spectrum was obtained, weighting individual subscans with the inverse square 
of their system temperature. The systemic local standard of rest velocity for
Maffei~2 is --14$\pm$5\,km\,s$^{-1}$ and $V_{\rm HEL}$ = $V_{\rm LSR}$ -- 3.1\,km\,s$^{-1}$.
For IC~342, the corresponding values are +35$\pm$3\,km\,s$^{-1}$ and $V_{\rm HEL}$ = 
$V_{\rm LSR}$ -- 3.9\,km\,s$^{-1}$. Vertical lines indicating the systemic 
velocities are included in the lower parts of the averaged spectra.}
\label{detections}
\end{figure*}

\subsection{Calibration}

We established flux density scales by performing continuum 
cross scans through 3C~286 and NGC~7027 (see Ott et al. 1994), also 
accounting for gain variations of the telescope as a function of elevation 
and for a 0.6\%\,yr$^{-1}$ secular decrease in intensity in the case of NGC~7027. 
Accounting for pointing errors and scatter in the measured signals, we
estimate that the calibration uncertainty is $\pm$15\%.

\section{Results}

Figure~\ref{detections} shows 36\,GHz methanol spectra taken towards the 
centers of Maffei~2 and IC~342. Following initial detections in early 
January 2019, which looked very promising but not fully convincing, 
we reobserved the sources a month later. With profiles from a total of 
three epochs, each showing signals above 3$\sigma$ and a total of order
10$\sigma$ (see Table~1), there is no doubt that the detections are real. 
However, we note that the lines obtained at the individual epochs are 
weak and relatively noisy. Therefore, differences with respect to peak flux 
density and line shape cannot be reliably interpreted in terms of variability. 
Instead, the differences are more likely caused by slightly varying average 
pointing offsets and, most likely, by different weather conditions, which 
included occasional light rain during the last observing night. This is 
consistent with the fact that the best single epoch spectra are from 
January 2019, while the latest spectra show the lowest signal-to-noise 
ratios.

Table~1 provides Gaussian fit parameters to the average spectra from 
Maffei~2 and IC~342. In Maffei~2 we find two well-separated spectral
components, while IC~342 exhibits one component and overall weaker 
emission, with lower peak and velocity integrated flux density. The 
CH$_3$OH line luminosities in the last column of Table~1, assuming
isotropic radiation, were calculated using 
\begin{equation}
   L/{\rm L_{\odot}} = 0.038 \times\ (S_{\nu}\,\Delta V_{1/2}/{\rm Jy\,km\,s^{-1}}) 
   \times\ (D/{\rm Mpc})^2, 
\end{equation}
with $S_{\nu}$, $\Delta V_{1/2}$ and $D$ representing flux density, full width
to half power line width and distance. Upper limits are provided for four 
additional galaxies as well as also for the 44\,GHz line of methanol towards 
Maffei~2 and IC~342.

\section{Discussion}

\subsection{Detected lines}

The two originally detected 36\,GHz galaxies outside the Local Group, NGC~253 
and NGC~4945 (Ellingsen et al. 2014; McCarthy et al. 2017) show overall 
properties that are quite different from the galaxies  detected by 
us. Both NGC~253 and NGC~4945 host ongoing starbursts and are characterized 
by infrared luminosities a few times higher than those of the Milky Way, 
Maffei~2 or IC~342. NGC~4945 also hosts an active galactic nucleus. NGC~253 
and NGC~4945 are both highly inclined giant spiral galaxies ($i$ $\approx$ 
75$^{\circ}$; e.g., Iodice et al. 2014; Henkel et al. 2018b). Furthermore, 
with outflowing gas in their nuclear regions (Turner 1985; Bolatto et al. 2013; 
Henkel et al. 2018b), they also provide an environment in which shocks potentially 
leading to class~I methanol maser emission may be ubiquitous. It 
is remarkable, that the various maser components in NGC~253 occur not right 
at the center but at offsets of a few hundred parsecs from the nucleus (Ellingsen et 
al. 2017). The same also holds in the case of NGC~4945, where the (southeastern) 
front side of the galaxy shows a single but relatively strong maser spot, 
presumably also arising at a distance of a few hundred parsecs from the nucleus 
(McCarthy et al. 2018b).

Maffei~2 and IC~342, located behind the Galactic plane, are different. 
To evaluate their global properties, the first notable parameter is distance. 
While IC~342 is located at about the same distance as NGC~253 and NGC~4945 
(here we adopt 3.28\,Mpc; Karachentsev et al. 2013), Maffei~2 may be located at 
a significantly greater distance. Wu et al. (2014) estimated distances
of 3.4 -- 3.5\,Mpc for both Maffei~2 and IC~342 using the tip of the red giant 
branch method.  However, Tikhonov \& Galazutdinova (2018) re-evaluated their 
Hubble Space Telescope ({\it HST}) images. They applied stellar photometry 
at infrared and optical spectral ranges and compared the resulting 
Hertzprung-Russel diagrams. In particular the red giant branches, with 
their upper luminosity boundaries and color indices, were analyzed 
and compared with data from galaxies being much less obscured by Galactic 
dust. These latter authors obtained extinctions of $A_{\rm v}$ = 3\ffm65 
and 0\ffm9 for Maffei~2 and IC~342, respectively. While their resulting 
distance to IC~342 is consistent with previously obtained values, their 
distance to Maffei~2 becomes 6.7$\pm$0.5\,Mpc. Reanalyzing near infrared 
photometry from the {\it HST}, Anand et al. (2019) confirm a larger distance 
for Maffei~2 and propose $D$ = 5.73$\pm$0.40\,Mpc, which is adopted in the 
following (see also Table~1 for both Maffei~2 and IC~342).

A common feature of Maffei~2 and IC~342 is, as already briefly mentioned, their 
location behind the plane of the Milky Way. While IC~342 is displaced from the 
plane by 10.$\!\!^{\circ}$6, Maffei~2 is almost exactly on this plane ($b^{\rm II}$ 
= --0.$\!\!^{\circ}$3), which led to its very late discovery (Maffei 
1968). With respect to infrared luminosity and therefore also to star formation, 
making use of the more recent distance estimates to Maffei~2, both galaxies 
have luminosities of $L_{\rm IR}$ $\approx$10$^{10}$\,L$_{\odot}$. Not 
hosting an AGN, they are about as active as our own Galaxy ($L_{\rm FIR}$ 
$\approx$ 1.2 $\times$ 10$^{10}$\,L$_{\odot}$; e.g., Misiriotis et al. 2004, 
2006), forming on average a few solar masses of new stars per year (see, 
e.g., Kennicutt (1998) for the correlation between infrared luminosity and star 
formation rate and Chomiuk \& Povich (2011) for uncertainties related to estimates 
of Galactic star formation rates). Both galaxies host two spiral arms 
terminating on a central ring with vigorous massive star formation 
(e.g., Meier \& Turner 2012). Therefore, chemically, Maffei~2 may be considered 
to some extent as a more inclined copy of IC~342 and may not (as before) 
be considered as a particularly small spiral galaxy. However, one major 
difference remains: it is the kiloparsec-sized bar in Maffei~2 that is 
possibly causing widespread shocks favoring 36\,GHz methanol maser emission 
(see Ellingsen et al. 2017 and below). A similar structure is not 
found in IC~342.

\begin{table*} \caption[]{Methanol Line parameters$^{\rm a, b, c, d}$} 
\begin{flushleft}
\begin{tabular}{llcccrrrcc}
\hline
Source          & $\alpha_{J2000}$ & $\delta_{J2000}$ &      Transition            & $\nu$   &  $\int{S}{\rm d}V$   & $V_{\rm LSR}$   & $\Delta V_{1/2}$& $S_{\rm peak}$ & $L_{\rm CH_3OH}$\\
                &                  &                  &                            &  GHz    &  mJy\,km\,s$^{-1}$   & km\,s$^{-1}$    & km\,s$^{-1}$    &    mJy         &  L$_{\odot}$   \\
\hline         
                &                  &                  &                            &         &                      &                 &                 &                &                 \\
Maffei~2        &    02:41:55.2    &     +59:36:12    & 4$_{-1}$$\rightarrow$3$_0$ &36.169261&    422$\pm$45        & --99.3$\pm$1.9  & 35.0$\pm$3.9    &   11.3$\pm$1.9 & 0.53$\pm$0.06   \\
                &                  &                  & 4$_{-1}$$\rightarrow$3$_0$ &36.169261&   1181$\pm$60        &    7.5$\pm$1.5  & 61.1$\pm$3.4    &   18.1$\pm$1.5 & 1.47$\pm$0.07   \\
                &                  &                  &  7$_0$$\rightarrow$6$_1$   &44.069419&                      &                 &                 &     $<$12.0    &                 \\
IC~342          &    03:46:48.6    &     +68:05:46    & 4$_{-1}$$\rightarrow$3$_0$ &36.169261&    641$\pm$64        &   29.7$\pm$3.6  & 70.6$\pm$8.0    &    8.5$\pm$1.4 & 0.26$\pm$0.03   \\
                &                  &                  &  7$_0$$\rightarrow$6$_1$   &44.069419&                      &                 &                 &     $<$10.0    &                 \\
M~82            &    09:55:52.2    &     +69:40:47    & 4$_{-1}$$\rightarrow$3$_0$ &36.169261&                      &                 &                 &      $<$4.4    &                 \\
NGC~4388        &    12:25:46.7    &     +12:39:41    & 4$_{-1}$$\rightarrow$3$_0$ &36.169261&                      &                 &                 &      $<$4.6    &                 \\
NGC~5728        &    14:42:23.9    &    --17:15:11    & 4$_{-1}$$\rightarrow$3$_0$ &36.169261&                      &                 &                 &      $<$4.5    &                 \\
Arp~220         &    15:34:57.3    &     +23:30:11    & 4$_{-1}$$\rightarrow$3$_0$ &36.169261&                      &                 &                 &      $<$4.2    &                 \\
\hline
\end{tabular}
\end{flushleft}
a) Obtained from Gaussian fits to the average spectra displayed in Fig.~\ref{detections}. 
Col.\,1: Sources; cols.\,2 and 3: $J$2000 coordinates; col.\,4: methanol transition; col.\,5: 
rest frequency; cols.\,6--8: integrated flux densities, local standard of rest velocities, and 
full width at half power line widths. Given errors include standard deviations from Gaussian fits 
but do not include the calibration uncertainty (Sect.\,2.2); col.\,9: peak flux densities; 
col.\,10: methanol line luminosities. Adopted distances: 5.73\,Mpc for Maffei~2 and 3.28\,Mpc 
for IC~342 (see Sect.\,4). \\
b) For M~82, NGC~4388, NGC~5728, and Arp~220, 1$\sigma$ upper limits are given for 9.7\,km\,s$^{-1}$ wide channels. \\
c) For the 44\,GHz transition towards Maffei~2 and IC~342, 1$\sigma$ upper limits are given for 8.0\,km\,s$^{-1}$ wide channels. \\
d) To convert the local standard of rest velocities given here into heliocentric values, use $V_{\rm HEL}$ = 
$V_{\rm LSR}$ -- 3.1\,km\,s$^{-1}$ for Maffei~2 and $V_{\rm HEL}$ = $V_{\rm LSR}$ -- 3.9\,km\,s$^{-1}$ for IC~342. 
\label{tab-B.1}
\end{table*}

The previously obtained 36\,GHz methanol maser detections in NGC~253 and 
NGC~4945 (Ellingsen et al. 2014; McCarthy et al. 2017) have something in 
common: the detected profiles are much narrower than those obtained when
measuring quasi-thermal line emission from areas covering their 
entire central molecular zones. This is different with respect to our 
lines shown in Fig.\,\ref{detections}. Galactic foreground emission 
(Maffei~2 and IC~342 are located behind the Perseus arm) cannot cause 
this effect. The two line peaks detected towards Maffei~2 are not 
compatible with molecular velocities of the Perseus arm (Cohen et al.
1980) and must therefore be of extragalactic origin. The 36\,GHz 
methanol velocity components are also seen, with similar spectral 
profiles but with lower signal-to-noise ratios, in other tracers 
of dense molecular gas, the integrated ammonia (NH$_3$) ($J,K$) 
= (1,1) and (likely) (2,2) profiles near 24\,GHz (Lebr{\'o}n et 
al. 2011). The global ammonia profiles from IC~342 also show 
similar shapes and radial velocities to those presented in 
Fig.~\ref{detections} and cannot be interpreted in terms of Galactic 
origin, since Galactic molecular clouds in this region show smaller 
radial velocities (Cohen et al. 1980; Lo et al. 1984; Tarchi et al. 2002). 
Both 36\,GHz methanol (e.g., Leurini et al. 2016) and 24\,GHz 
ammonia (e.g., Lebr{\'o}n et al. 2011) may trace similarly dense molecular 
gas because rates for spontaneous emission are not far apart (1.48 
and 1.67 $\times$ 10$^{-7}$\,s$^{-1}$, the latter for the ($J,K)$ = (1,1) 
ammonia transition; e.g., Sch{\"o}ier et al. 2005 and references
therein). Therefore, the agreement in line shapes does not directly 
hint at 36\,GHz maser emission in our two detected galaxies, but this 
is further discussed below in relation to what is known from the 
central region of the Milky Way.

\begin{figure}[ht]
\hspace{0.3cm}
\resizebox{13.3cm}{!}{\rotatebox[origin=br]{-90}{\includegraphics{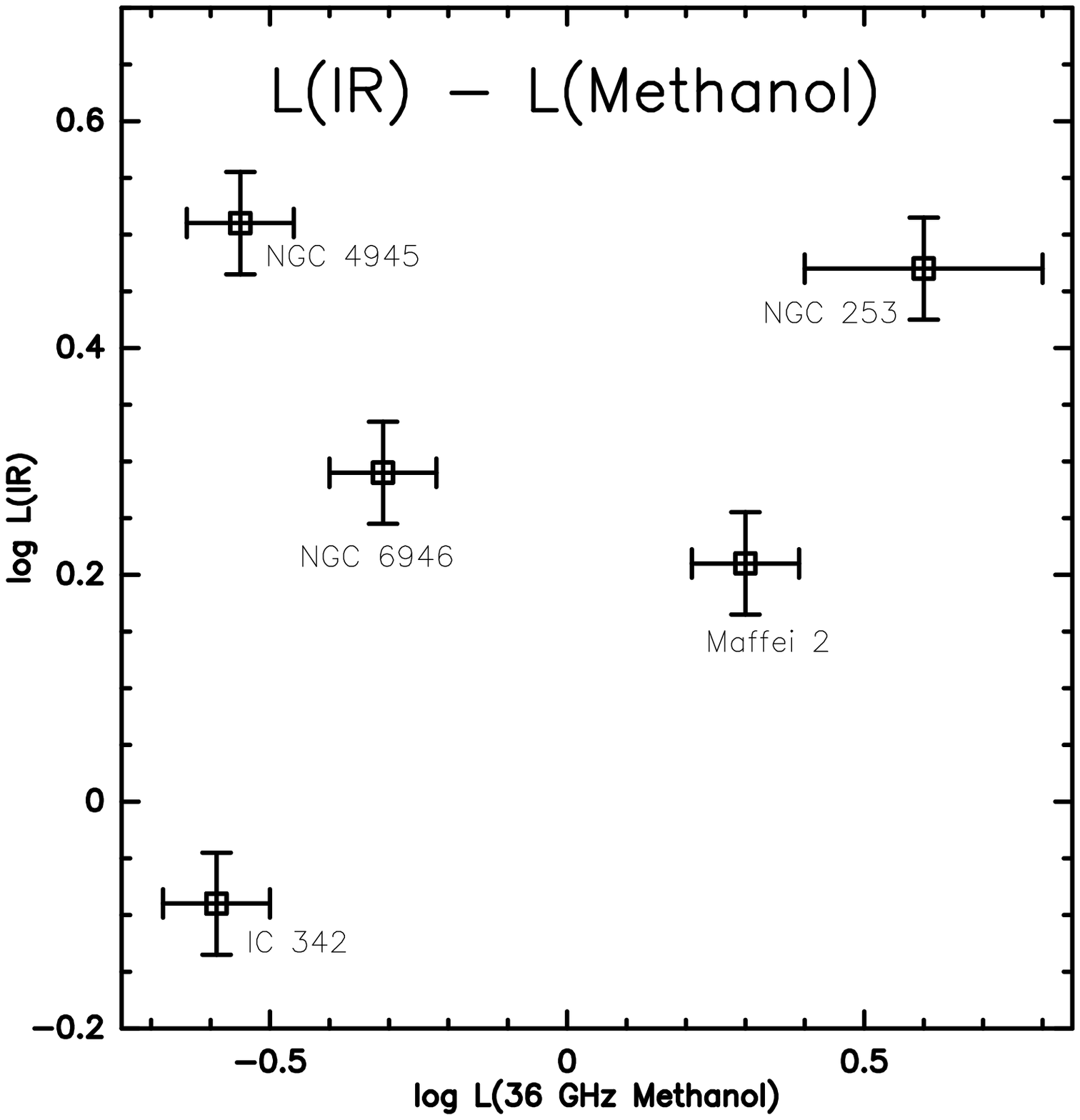}}}
\vspace{0.0cm}
\caption{Here we show 36\,GHz methanol line luminosities in solar units vs. 
infrared luminosities in units of 10$^{10}$\,L$_{\odot}$, obtained from IRAS 
data, on logarithmic scales. For the former, $\pm$20\% (see Sect.\,2.2, we 
added 5\% due to limited signal-to-noise ratios) and for the latter $\pm$10\% 
error bars have been taken. The exception is NGC~253, where a larger error 
bar has been chosen for the 36\,GHz methanol luminosity (see below). Adopted 
distances to the galaxies are 3.28\,Mpc for IC~342 (Sect.\,4), 5.73\,Mpc for 
Maffei~2 (Sect.\,4), 5.98\,Mpc for NGC~6946 (Gorski et al. 2018), 3.5\,Mpc 
for NGC~253 (Gorski et al. 2018) and 3.8\,Mpc for NGC~4945 (Henkel et al. 
2018b).  The methanol line luminosities were calculated with equation (1). 
There exist huge discrepancies in total 36\,GHz methanol flux density 
in the case of NGC~253, with Ellingsen et al. (2014) finding $\approx$0.3\,L$_{\odot}$, 
Ellingsen et al. (2017) reporting $\approx$3\,L$_{\odot}$ and Gorski et al. (2017) 
suggesting 6.4\,L$_{\odot}$. We chose 4\,L$_{\odot}$. For NGC~4945, McCarthy 
et al. (2017) report 0.14\,L$_{\odot}$, while McCarthy et al. (2018b) find 
2.8\,L$_{\odot}$. We have taken the latter value, suspecting that the lower 
luminosity values are due to missing flux (e.g., Chen et al. 2018) and that 
variability might also play a role.}
\label{ch3oh-ir}
\end{figure}

Following the nomenclature of Lebr{\'o}n et al. (2011), used for their ammonia 
(NH$_3$) observations with a beam size of $\approx$3\ffas5, the blueshifted component 
we recognize in Maffei~2 (Fig.~\ref{detections} and Table~1) must arise from their
regions C and D, located in the southern part of the galaxy. The component 
near zero velocity is related to the northern part, represented by 
molecular complexes A and B. In Fig.~2 of Lebr{\'o}n et al. (2011), sources 
B and C are dominant and, with their separation of 10$''$ -- 15$''$ (280\,pc --
415\,pc), they should also dominate the 36\,GHz methanol emission inside our 
23$''$ beam. In the para-NH$_3$ (1,1) ground state inversion transition, 
the redshifted component is stronger than the blueshifted one, by a factor of 
roughly 1.6. The corresponding line intensity ratio we see in our 36\,GHz 
methanol transition is 2.8$\pm$0.3 in favor of this more redshifted component. 
While this difference is clearly compatible with quasi-thermal emission in both 
NH$_3$ (where this is expected) and CH$_3$OH, although suggesting slightly 
different physical or chemical conditions, we nevertheless note that the blueshifted 
CH$_3$OH component may be narrower than the one seen in ammonia. 

Toward IC~342, our line widths and velocities of methanol are comparable,
within the errors, to the ammonia lines from Lebr{\'o}n et al. (2011). Very recently, 
Gorski et al. (2018) reported JVLA measurements of the central kiloparsec of IC~342
with 1$''$ and 7\,km\,s$^{-1}$ resolution. These latter authors  detected six spatially 
resolved 36\,GHz CH$_3$OH sites in IC~342 located along a northeast-southwest
axis (position angle $\approx$30$^{\circ}$) with a length of 20$''$ (300\,pc). 
Interestingly, their strongest four sources peak at velocities clearly 
redshifted ($\approx$50\,km\,s$^{-1}$) with respect to the galaxy's systemic 
local standard of rest velocity of 35\,km\,s$^{-1}$. A similar trend is 
also seen in the NH$_3$ (1,1) emission from Lebr{\'o}n et al. (2011; their 
Fig.~1, lower right panel). However, our Table~1 instead indicates a peak 
velocity of the overall emission at systemic or slightly blueshifted 
velocities. Since the Gorski et al. (2018) data encompass the entire 
central region of IC~342 giving rise to emission from high-density 
molecular tracers, this discrepancy cannot be explained in terms of spatial 
regions not being covered by their measurements. This also holds for our 
data. Missing flux in the interferometric measurements might instead be 
the cause of the difference. However, adding the luminosities given in 
their Table~6 leads to a total luminosity of 0.315\,L$_{\odot}$, 
which is (within 2$\sigma$, see, e.g., Sect.\,2.2) compatible 
with our luminosity of (0.26 $\pm$ 0.03)\,L$_{\odot}$ (Table~1). 
While we cannot explain the difference in radial velocities, 
the apparent lack of missing flux suggests that the bulk of the 36\,GHz 
methanol emission is originating from compact spatially unresolved regions, 
compatible with maser emission. 

In the folllowing text we use the definition of starburst galaxies as
outlined by Mao et al. (2010; their Fig.~1), where the optical isophotal 
diameter $D_{25}$ (25$^m$/arcsec$^2$) is applied to obtain the ratio 
log\,[($L_{\rm IR}$/L$_{\odot}$)/($D_{25}\!\!^2$/kpc$^{2}$)]. Here, in
almost all cases the value 7.25 separates galaxies commonly identified 
as starburst from nonstarburst galaxies. With Maffei~2, IC~342 and 
NGC~6946 not being starburst galaxies like NGC~253 and NGC~4945 
(according to our definition), but showing infrared luminosities 
and likely star forming rates similar to those of the Milky Way, 
there may be similarities between the conditions in Maffei~2, 
IC~342, and NGC~6946 and the central molecular zone (CMZ) of 
our Galaxy. Haschick et al. (1990) provided data from our Galactic 
center region, showing 36\,GHz and 44\,GHz methanol spectra from 
Sgr~A-A $\approx$1$'$ northeast of the Galactic center and Sgr~A-F, 
$\approx$3$'$ south of the Galactic center (their Fig.~9). These latter 
authors find narrow 36\,GHz spikes, likely due to maser emission, on top of a 
broad component, possibly of quasi-thermal origin. For these two Galactic 
sources, the 36\,GHz methanol lines show quite a number of narrow spikes, 
while only one such spike is seen in the corresponding 44\,GHz methanol line. 

More recently, Yusef-Zadeh et al. (2013) and Cotton \& Yusef-Zadeh (2016) 
find a widespread population of more than 2000 compact 36\,GHz methanol 
maser sources within the central degree of our Galaxy. The lines are narrow 
($\approx$1\,km\,s$^{-1}$). Such a widespread distribution with a huge 
number of individual maser components covering the full radial velocity 
range of the nuclear environment could mimic a total profile, that is quite 
undistinguishable from quasi-thermal emission as long as the masers
are spatially unresolved and no individual source stands out. 

Figure~\ref{ch3oh-ir} correlates 36\,GHz line methanol luminosities 
with infrared luminosities obtained from IRAS (the Infrared Astronomical 
Satellite) and therefore with the star forming rates of their respective 
parent galaxies. While the number of sources is small and while both 
36\,GHz line and infrared continuum luminosities may be uncertain, it 
is clear that there is no good correlation. NGC~253 with its high infrared 
luminosity shows the most luminous 36\,GHz methanol emission. However,
NGC~4945, almost as luminous in the infrared, is comparatively weak
with respect to the 36\,GHz methanol line. Outflowing gas in its nuclear 
region (Henkel et al. 2018b), leading to shocks, is possibly not quite 
as prominent as in NGC~253 (Turner 1985; Bolatto et al. 2013). Maffei~2 
also stands out with respect to methanol emission, even though its 
infrared luminosity is modest. While we did not resolve the sources, 
the explanation may be the prominent nuclear bar in Maffei~2 (e.g., 
Meier \& Turner 2012), that may cause shocks throughout its central 
region, enhancing 36\,GHz methanol emission above its normal level. 
Higher-resolution observations to study the spatial connection between 
the bar and the 36\,GHz methanol emission would be desirable. 

As already mentioned, it is noteworthy that toward the nonstarbursting 
galaxies we see line profiles compatible with the line profiles of other dense 
molecular gas tracers (for NH$_3$, see the discussion above; for HNCO, see 
Gorski et al. 2017, 2018). Toward the two detected starburst galaxies,
however, we see emission from highly confined regions, the strongest 
such region being located in NGC~4945. We may therefore ask whether these 
galaxies also contain a component similar to the one in the less
active galaxies Maffei~2, IC~342, and NGC~6946, only being less
conspicuous in view of the outstanding maser peaks already encountered 
in these starburst environments. We advocate searches for such 
an underlying smoother 36\,GHz emission component also in these 
galaxies. We note however that by analogy with our Galactic center 
region (e.g., Yusef-Zadeh et al. 2013), it is likely that this 
component is also representing maser emission.

\subsection{Undetected lines}

It may be argued that our lack of detected 44\,GHz methanol in Maffei~2 and 
IC~342 is caused by the different levels of excitation, with the 36\,GHz upper 
level (4$_{-1}$) 21\,K and the 44\,GHz upper level (7$_0$) 65\,K above the ground 
state (the equivalent energy of the photons is only $\approx$2\,K). However, we 
note that our 44\,GHz measurements show rms values that are about three times 
larger than those at 36\,GHz. Therefore, our nondetections at 44\,GHz toward
Maffei~2 and IC~342 merely indicate that the 44\,GHz emission is not stronger
than that at 36\,GHz, which is consistent with the conditions in NGC~253
(Ellingsen et al. 2017) and appears to be a typical signature of 
extragalactic methanol class~I emission. 

With respect to our 36\,GHz nondetections, the Arp~220 result is particularly
interesting. Chen et al. (2015) reported several areas with emission in the 
outskirts of the merger with flux densities of order 10--20\,mJy and our 
beam size is large enough to incorporate most of these regions. However,
our upper limit given in Table~1 (last line) corresponds to a 5$\sigma$ level 
of $\approx$4\,mJy for a feature of 300\,km\,s$^{-1}$ in width. In view of 
its extreme infrared flux density, M~82 was an obvious target (it is one of 
the most prolific extragalactic infrared emitters of the entire sky; e.g., 
Henkel et al. 1986). Instead, NGC~4388 and NGC~5728 were observed because 
Chen et al. (2016) had published tentative 36\,GHz class~I methanol signals. 
We were not able to confirm their tentative features but note that our data 
are of similar sensitivity, and so in these cases, unlike for Arp~220,
the situation remains unclear.

\section{Conclusions}

With Maffei~2 and IC~342 we present two 36\,GHz class~I methanol 
line emitters from galaxies outside the Local Group, while a previously
reported detection of the ultraluminous infrared galaxy Arp~220 is not 
confirmed. The line luminosity of Maffei~2 is surprisingly high, which 
is possibly related to its prominent bar. As in the prototypical source, 
NGC~253, upper limits indicate that the 44\,GHz class~I methanol line is 
not substantially stronger. Now there are five such well-confirmed sources 
with 36\,GHz line detections beyond any reasonable doubt. In contrast to 
the two initially detected and still most prominent of these sources, 
NGC~253 and NGC~4945, it is difficult to directly prove that the detected 
signals from Maffei~2 and IC~342 (and NGC~6946) are caused by masers. 
The line profiles are compatible with those obtained from ammonia, where 
masers are rarely observed. Unlike NGC~253 and NGC~4945, which are starburst 
galaxies, the others are spiral galaxies that are about as active with respect 
to star formation as our Milky Way.  Therefore, by analogy with the central 
part of our Galaxy, the most likely interpretation is that the detected 
features represent the superposition of a large number of spatially 
unresolved weak maser hotspots lacking a single dominant source. It 
is hard to imagine that more active galaxies, like for example NGC~253 and 
NGC~4945, do not also show such a population of weak masers. Therefore, 
a search for such a weak and broad underlying spatially widespread 
36\,GHz methanol emission would also be worthwhile in starburst 
galaxies, as well as interferometric studies with utmost sensitivity to 
unambiguously demonstrate that at least a part of the detected emission originates 
from maser sources. The latter would, however, be a very ambitious project,
presumably only feasible with the future next generation Very Large Array (ngVLA).

\acknowledgements
We wish to thank K.-M. Menten and an anonymous referee for valuable comments 
as well as the staff at Effelsberg for their great support, in particular the operators 
and A. Kraus. S.A.L. acknowledges funding through RSF grant No. 19-12-00157. This 
research has made use of NASA's Astrophysical Data System and of the NASA/IPAC 
Extragalactic Database.

\end{document}